\pdfoutput=1
\documentclass[12pt,a4paper]{article}

\usepackage{ifthen} 
\newboolean{pdflatex}
\setboolean{pdflatex}{true} 

\newboolean{articletitles}
\setboolean{articletitles}{true} 

\newboolean{uprightparticles}
\setboolean{uprightparticles}{false} 

\newboolean{inbibliography}
\setboolean{inbibliography}{false} 

\def\paperauthors{The Heavy Flavor Averaging Group (HFLAV)} 
\def\paperasciititle{HFLAV input to the 2026 update of the European Strategy for Particle Physics} 
\def\papertitle{HFLAV input to the 2026 update of the \\ European Strategy for Particle Physics} 
\def\paperkeywords{{High Energy Physics}, {HFLAV}, {flavour physics}} 
\def\papercopyright{2025 HFLAV} 
\def\paperlicence{CC-BY-4.0 licence}
\def\paperlicenceurl{https://creativecommons.org/licenses/by/4.0/}

\usepackage[top=1in, bottom=1.25in, left=1in, right=1in]{geometry}

\usepackage[utf8]{inputenc}
\usepackage[T1]{fontenc}
\usepackage{microtype}
\usepackage{lineno}  
\usepackage{xspace} 
\usepackage{caption} 

\usepackage{graphicx}  
\usepackage{color}
\usepackage{colortbl}
\graphicspath{{./figs/}} 
\DeclareGraphicsExtensions{.pdf,.PDF,png,.PNG}

\usepackage{amsmath} 
\usepackage{amssymb}
\usepackage{amsfonts}
\usepackage{upgreek} 

\newcommand*\patchAmsMathEnvironmentForLineno[1]{%
\expandafter\let\csname old#1\expandafter\endcsname\csname #1\endcsname
\expandafter\let\csname oldend#1\expandafter\endcsname\csname
end#1\endcsname
 \renewenvironment{#1}%
   {\linenomath\csname old#1\endcsname}%
   {\csname oldend#1\endcsname\endlinenomath}%
}
\newcommand*\patchBothAmsMathEnvironmentsForLineno[1]{%
  \patchAmsMathEnvironmentForLineno{#1}%
  \patchAmsMathEnvironmentForLineno{#1*}%
}
\AtBeginDocument{%
\patchBothAmsMathEnvironmentsForLineno{equation}%
\patchBothAmsMathEnvironmentsForLineno{align}%
\patchBothAmsMathEnvironmentsForLineno{flalign}%
\patchBothAmsMathEnvironmentsForLineno{alignat}%
\patchBothAmsMathEnvironmentsForLineno{gather}%
\patchBothAmsMathEnvironmentsForLineno{multline}%
\patchBothAmsMathEnvironmentsForLineno{eqnarray}%
}

\usepackage[pdftex,
            pdfauthor={\paperauthors},
            pdftitle={\paperasciititle},
            pdfkeywords={\paperkeywords}]{hyperref}
\usepackage{hyperxmp}

\usepackage[colorinlistoftodos,textsize=scriptsize]{todonotes}
\usepackage{comment}

\usepackage[all]{hypcap} 
\usepackage{authblk}
\usepackage{orcidlink}  

\usepackage{xspace} 
\usepackage{upgreek}

\newcommand{\offsetoverline}[2][0.1em]{\kern #1\overline{\kern -#1 #2}}%

\def\lhcb   {\mbox{LHCb}\xspace}
\def\atlas  {\mbox{ATLAS}\xspace}
\def\cms    {\mbox{CMS}\xspace}

\def\babar  {\mbox{BaBar}\xspace}
\def\belle  {\mbox{Belle}\xspace}
\def\belletwo {\mbox{Belle~II}\xspace}
\def\besiii {\mbox{BESIII}\xspace}

\def\MagUp {\mbox{\em Mag\kern -0.05em Up}\xspace}

\ifthenelse{\boolean{uprightparticles}}%
{

 \def\PDelta      {\ensuremath{\Delta}\xspace}                 
 \def\PXi         {\ensuremath{\Xi}\xspace}                 
 \def\PLambda     {\ensuremath{\Lambda}\xspace}                 
 \def\PSigma      {\ensuremath{\Sigma}\xspace}                 
 \def\POmega      {\ensuremath{\Omega}\xspace}                 
 \def\PUpsilon    {\ensuremath{\Upsilon}\xspace}

 \def\PB      {\ensuremath{\mathrm{B}}\xspace}                 
                  
 \def\PD      {\ensuremath{\mathrm{D}}\xspace}

 \def\PK      {\ensuremath{\mathrm{K}}\xspace}

 \def\Pc      {\ensuremath{\mathrm{c}}\xspace}                 
                  
 \def\Pe      {\ensuremath{\mathrm{e}}\xspace}

 \def\Pi      {\ensuremath{\mathrm{i}}\xspace}

}
{

 \mathchardef\PDelta="7101
 \mathchardef\PXi="7104
 \mathchardef\PLambda="7103
 \mathchardef\PSigma="7106
 \mathchardef\POmega="710A
 \mathchardef\PUpsilon="7107
                  
 \def\PB      {\ensuremath{B}\xspace}                 
                  
 \def\PD      {\ensuremath{D}\xspace}

 \def\PK      {\ensuremath{K}\xspace}

 \def\Pc      {\ensuremath{c}\xspace}                 
                  
 \def\Pe      {\ensuremath{e}\xspace}

 \def\Pi      {\ensuremath{i}\xspace}

}

\makeatletter
\ifcase \@ptsize \relax
  \newcommand{\miniscule}{\@setfontsize\miniscule{4}{5}}
\or
  \newcommand{\miniscule}{\@setfontsize\miniscule{5}{6}}
\or
  \newcommand{\miniscule}{\@setfontsize\miniscule{5}{6}}
\fi
\makeatother

\DeclareRobustCommand{\optbar}[1]{\shortstack{{\miniscule (\rule[.5ex]{1.25em}{.18mm})}
  \\ [-.7ex] $#1$}}

\def\epem       {{\ensuremath{\Pe^+\Pe^-}}\xspace}

\def\cquark    {{\ensuremath{\Pc}}\xspace}
\def\cquarkbar {{\ensuremath{\overline \cquark}}\xspace}

  \def\Kbar    {{\kern 0.2em\overline{\kern -0.2em \PK}{}}\xspace}

\def\KorKbar {\kern 0.18em\optbar{\kern -0.18em K}{}\xspace}

  \def\Dbar    {{\kern 0.2em\overline{\kern -0.2em \PD}{}}\xspace}

\def\DorDbar {\kern 0.18em\optbar{\kern -0.18em D}{}\xspace}

\def\Bbar    {{\ensuremath{\kern 0.18em\overline{\kern -0.18em \PB}{}}}\xspace}

\def\BorBbar    {\kern 0.18em\optbar{\kern -0.18em B}{}\xspace}

\def\Y#1S{\ensuremath{\PUpsilon{(#1S)}}\xspace}

\def\LorLbar     {\kern 0.18em\optbar{\kern -0.18em \PLambda}{}\xspace}

\def\to                 {\ensuremath{\rightarrow}\xspace}

\def\CP                {{\ensuremath{C\!P}}\xspace}

\def\AT#1     {\ensuremath{A_{\mathrm{T}}^{#1}}\xspace}           

\def\C#1      {\ensuremath{\mathcal{C}_{#1}}\xspace}                       
\def\Cp#1     {\ensuremath{\mathcal{C}_{#1}^{'}}\xspace}                    
\def\Ceff#1   {\ensuremath{\mathcal{C}_{#1}^{\mathrm{(eff)}}}\xspace}        
\def\Cpeff#1  {\ensuremath{\mathcal{C}_{#1}^{'\mathrm{(eff)}}}\xspace}       
\def\Ope#1    {\ensuremath{\mathcal{O}_{#1}}\xspace}                       
\def\Opep#1   {\ensuremath{\mathcal{O}_{#1}^{'}}\xspace}                    

\newcommand{\tev}{\ifthenelse{\boolean{inbibliography}}{\ensuremath{~T\kern -0.05em eV}}{\ensuremath{\mathrm{\,Te\kern -0.1em V}}}\xspace}
\newcommand{\gev}{\ensuremath{\mathrm{\,Ge\kern -0.1em V}}\xspace}
\newcommand{\mev}{\ensuremath{\mathrm{\,Me\kern -0.1em V}}\xspace}
\newcommand{\kev}{\ensuremath{\mathrm{\,ke\kern -0.1em V}}\xspace}
\newcommand{\ev}{\ensuremath{\mathrm{\,e\kern -0.1em V}}\xspace}
\newcommand{\mevc}{\ensuremath{{\mathrm{\,Me\kern -0.1em V\!/}c}}\xspace}
\newcommand{\gevc}{\ensuremath{{\mathrm{\,Ge\kern -0.1em V\!/}c}}\xspace}
\newcommand{\mevcc}{\ensuremath{{\mathrm{\,Me\kern -0.1em V\!/}c^2}}\xspace}
\newcommand{\gevcc}{\ensuremath{{\mathrm{\,Ge\kern -0.1em V\!/}c^2}}\xspace}
\newcommand{\gevgevcc}{\ensuremath{{\mathrm{\,Ge\kern -0.1em V^2\!/}c^2}}\xspace} 
\newcommand{\gevgevcccc}{\ensuremath{{\mathrm{\,Ge\kern -0.1em V^2\!/}c^4}}\xspace} 

\def\gsim{{~\raise.15em\hbox{$>$}\kern-.85em
          \lower.35em\hbox{$\sim$}~}\xspace}
\def\lsim{{~\raise.15em\hbox{$<$}\kern-.85em
          \lower.35em\hbox{$\sim$}~}\xspace}

\def\tell1  {TELL1\xspace}
\def\ukl1   {UKL1\xspace}

\newcommand{\eg}{\mbox{\itshape e.g.}\xspace}

\usepackage{cite} 
\usepackage{mciteplus}

\usepackage{bm}

\begin{document}

\renewcommand\Affilfont{\itshape\small}

\author[1]{F.~Archilli\,\orcidlink{0000-0002-1779-6813}}\affil[1]{INFN Sezione di Roma Tor Vergata and Universit\`a di Roma Tor Vergata, Roma, Italy}
\author[2]{Sw.~Banerjee\,\orcidlink{0000-0001-8852-2409}}\affil[2]{University of Louisville, Louisville, Kentucky, USA}
\author[3]{E.~Ben-Haim\,\orcidlink{0000-0002-9510-8414}}\affil[3]{LPNHE, Sorbonne Universit\'e, Paris Diderot Sorbonne Paris Cit\'e, CNRS/IN2P3, Paris, France}
\author[4]{F.~U.~Bernlochner\,\orcidlink{0000-0001-8153-2719}}\affil[4]{University of Bonn, Bonn, Germany}
\author[5]{E.~Bertholet\,\orcidlink{0000-0002-3792-2450}}\affil[5]{Tel Aviv University, Tel Aviv, Israel}
\author[6]{M.~Bona\,\orcidlink{0000-0002-9660-580X}}\affil[6]{Department of Physics and Astronomy, Queen Mary University of London, London, UK}
\author[7]{A.~Bozek\,\orcidlink{0000-0002-5915-1319}}\affil[7]{H. Niewodniczanski Institute of Nuclear Physics, Krak\'{o}w, Poland}
\author[8]{C.~Bozzi\,\orcidlink{0000-0001-6782-3982}}\affil[8]{INFN Sezione di Ferrara, Ferrara, Italy}
\author[7]{J.~Brodzicka\,\orcidlink{0000-0002-8556-0597}}
\author[9]{V.~Chobanova\,\orcidlink{0000-0002-1353-6002}}\affil[9]{Universidade da Coruna, La Coruna, Spain}
\author[7]{M.~Chrzaszcz\,\orcidlink{0000-0001-7901-8710}}
\author[10]{M. Dorigo\,\orcidlink{0000-0002-0681-6946}}\affil[10]{INFN Sezione di Trieste, Trieste, Italy}
\author[11]{U.~Egede\,\orcidlink{0000-0001-5493-0762}}\affil[11]{School of Physics and Astronomy, Monash University, Melbourne, Australia}
\author[12]{A.~Gaz\,\orcidlink{0000-0001-6754-3315}}\affil[12]{INFN Sezione di Padova and Universit\`a degli Studi di Padova, Padova, Italy}
\author[13]{M.~Gersabeck\,\orcidlink{0000-0002-0075-8669}}\affil[13]{Physikalisches Institut, Albert-Ludwigs-Universität Freiburg, Freiburg im Breisgau, Germany}
\author[14]{T.~Gershon\,\orcidlink{0000-0002-3183-5065}}\affil[14]{Department of Physics, University of Warwick, Coventry, United Kingdom}
\author[15]{P.~Goldenzweig\,\orcidlink{0000-0001-8785-847X}}\affil[15]{Institut f\"ur Experimentelle Teilchenphysik, Karlsruher Institut f\"ur Technologie, Karlsruhe, Germany}
\author[16]{L. Grillo\,\orcidlink{0000-0001-5360-0091}}\affil[16]{University of Glasgow, Glasgow, United Kingdom}
\author[17]{K.~Hayasaka\,\orcidlink{0000-0002-6347-433X}}\affil[17]{Niigata University, Niigata, Japan}
\author[18]{T.~Humair\,\orcidlink{0000-0002-2922-9779}}\affil[18]{Deutsches Elektronen-Synchrotron (DESY), Hamburg, Germany}
\author[19]{D.~Johnson\,\orcidlink{0000-0003-3272-6001}}\affil[19]{School of Physics and Astronomy, University of Birmingham, Birmingham, United Kingdom}
\author[20]{M.~Kenzie\,\orcidlink{0000-0001-7910-4109}}\affil[20]{Cavendish Laboratory, University of Cambridge, Cambridge, United Kingdom}
\author[21]{T.~Kuhr\,\orcidlink{0000-0001-6251-8049}}\affil[21]{Ludwig-Maximilians-University, Munich, Germany}
\author[22]{O.~Leroy\,\orcidlink{0000-0002-2589-240X}}\affil[22]{Aix Marseille Univ, CNRS/IN2P3, CPPM, Marseille, France}
\author[23]{A.~Lusiani\,\orcidlink{0000-0002-6876-3288}}\affil[23]{INFN Sezione di Pisa and Scuola Normale Superiore, Pisa, Italy}
\author[24]{H.-L.~Ma\,\orcidlink{0000-0001-9771-2802}}\affil[24]{Institute of High Energy Physics, Beijing 100049, People's Republic of China}
\author[12]{M.~Margoni\,\orcidlink{0000-0003-4352-734X}}
\author[25]{R.~Mizuk\,\orcidlink{0000-0002-2209-6969}}\affil[25]{Paris-Saclay University, CNRS/IN2P3, IJCLab, Orsay, France}
\author[26]{P.~Naik\,\orcidlink{0000-0001-6977-2971}}\affil[26]{Oliver Lodge Laboratory, University of Liverpool, Liverpool, United Kingdom}
\author[4]{M.~T.~Prim\,\orcidlink{0000-0002-1407-7450}}
\author[27]{M.~Roney\,\orcidlink{0000-0001-7802-4617}}\affil[27]{University of Victoria, Victoria, British Columbia, Canada}
\author[28]{M.~Rotondo\,\orcidlink{0000-0001-5704-6163}}\affil[28]{Laboratori Nazionali dell'INFN di Frascati, Frascati, Italy}
\author[29]{O.~Schneider\,\orcidlink{0000-0002-6014-7552}}\affil[29]{Institute of Physics, Ecole Polytechnique F\'{e}d\'{e}rale de Lausanne (EPFL), Lausanne, Switzerland}
\author[30]{C.~Schwanda\,\orcidlink{0000-0003-4844-5028}}\affil[30]{Institute of High Energy Physics, Vienna, Austria}
\author[31]{A.~J.~Schwartz\,\orcidlink{0000-0002-7310-1983}}\affil[31]{University of Cincinnati, Cincinnati, Ohio, USA}
\author[22]{J.~Serrano\,\orcidlink{0000-0003-2489-7812}}
\author[32]{B.~Shwartz\,\orcidlink{0000-0002-1456-1496}}\affil[32]{Budker Institute of Nuclear Physics, Novosibirsk, Russia}
\author[33]{M.~Veronesi\,\orcidlink{0000-0002-1916-3884}}\affil[33]{Iowa State University, Ames, Iowa 50011, USA}
\author[16]{M.~Whitehead\,\orcidlink{0000-0002-2142-3673}}
\author[34]{J.~Yelton\,\orcidlink{0000-0001-8840-3346}}\affil[34]{University of Florida, Gainesville, Florida, USA}

\title{\papertitle \vspace{4mm}  \\ \large \paperauthors\\
\href{mailto:hflav-conveners@cern.ch}{hflav-conveners@cern.ch}}
\date{}

\begin{titlepage}


\vspace*{4cm}
\begin{center}
{\LARGE \bfseries \papertitle \par}
\vspace{1cm}   \large \paperauthors \\
\href{mailto:hflav-conveners@cern.ch}{hflav-conveners@cern.ch}   
\end{center}
\vspace{1cm}


\begin{abstract}
  \noindent 
    Heavy-flavour physics is an essential component of the particle-physics programme, offering critical tests of the Standard Model and far-reaching sensitivity to physics beyond it. Experiments such as \lhcb, \belletwo, and \besiii 
    drive progress in the field, along with contributions from ATLAS and CMS. The LHCb Upgrade II and upgraded Belle II experiments will provide unique and highly sensitive measurements for decades, playing a key role in the searches for new physics. Future facilities with significant heavy-flavour capabilities will further expand these opportunities. We advocate for a European Strategy that fully supports Upgrade II of LHCb and an upgrade of \belletwo, along with their subsequent exploitation. Additionally, we support a long-term plan that fully integrates flavour physics in an $\epem$ collider to run as a $Z$ factory.
\end{abstract}
\vfill
{\footnotesize 
\centerline{\copyright~\papercopyright. \href{\paperlicenceurl}{\paperlicence}.}}
\vspace*{2mm}

\renewcommand{\thefootnote}{\arabic{footnote}}
\setcounter{footnote}{0}
\end{titlepage}

\newpage
\begin{center}
\paperauthors\\
\vspace{8mm}
\makeatletter
\@author
\makeatother
\end{center}

\parskip=2mm

\newpage

\noindent 
The Heavy Flavor Averaging Group (HFLAV)~\cite{HFLAV-web} is an international collaboration of physicists from experiments measuring properties of beauty and charm hadrons as well as $\tau$ leptons.  
HFLAV calculates and publishes world averages of measurements from current and past experiments, and provides a comprehensive resource for the field in terms of web pages and full documentation of results. 
The most recent compilation of our results appears in Ref.~\cite{HFLAV24}. Many of our world averages are used by the Particle Data Group~\cite{PDG2024}. 
With this perspective, we take the opportunity to comment on the importance of future heavy-flavour physics research. 
\textbf{In this document, we advocate for strong support of heavy-flavour physics, both in the near and the far future.} 

Flavour physics is among the central foundations of the Standard Model
(SM). Indeed, many advances in the construction of the SM originated
from research into flavour physics. This includes the three-generation prediction of the
Kobayashi-Maskawa mechanism, the universality of the gauge interactions, the high
masses of the top quark and the weak gauge bosons, and the presence of
large charge-parity (\CP) violation in beauty hadrons. Similarly, heavy-flavour physics is
closely tied to questions that aim to unravel the physics beyond the SM
(BSM). Some of these questions, and the ways in which current research
in heavy-flavour physics addresses them, are as follows:
\begin{itemize}
\item The origins of the three generations of fermions and of the Yukawa couplings that distinguish between them are not explained by the SM. Testing at high precision the Cabibbo-Kobayashi-Maskawa (CKM) matrix might reveal that the three-generation picture is incomplete or that the quark-flavour nonuniversality does not depend only on the four parameters of the CKM matrix. 

\item The SM structure implies several accidental symmetries, such as 
    the universality of gauge interactions for charged leptons, 
    the conservation of charged-lepton flavour and lepton and baryon numbers. 
    Any violation of these properties is an unambiguous sign of BSM physics. 
    Similarly, testing flavour-changing neutral currents, which in the SM occur only at loop level, offers very sensitive probes to the presence of heavy new states.  

  \item The baryon asymmetry of the universe necessitates \CP violation 
  far beyond that provided by the SM. Precise measurements of \CP violation in heavy-flavour 
  decays may uncover new sources of \CP violation. 
  Processes in which the SM predicts zero or very small \CP violation 
  can be particularly sensitive to BSM amplitudes. These
  include decays of the $\tau$ lepton and specific charm- and
  beauty-hadron decays.
\end{itemize}

\noindent Heavy-flavour measurements lead to tight constraints on BSM physics, in many cases at energy scales that are far beyond
those accessible at energy-frontier facilities. Currently, some measurements suggest potential deviations from SM predictions, 
such as the averages of the lepton-universality ratios $R(D^{(*)})$, 
some branching fractions and angular observables in $b \to s \ell^+\ell^-$ transitions, 
or the recent evidence of the rare $B^+ \to K^+ \nu\bar\nu$ decay~\cite{HFLAV24,LHCb:2024onj,Belle-II:2023esi}.   
Additional data and independent confirmations among different experiments, as well as improvements in the theoretical understanding, are needed to establish any discrepancy conclusively.  
If these deviations stem from BSM physics, a consistent pattern of further discrepancies in heavy-flavour decays should emerge, paving the way for new model development.
Thus, \textbf{heavy-flavour physics has far-reaching BSM sensitivity, as well as an important role in informing research at current and future energy-frontier facilities.} 

In addition, a number of hadrons with nonstandard quantum numbers that contain charm or beauty quarks, 
(\eg, $\chi_{c 1}(3872)$, $T_{\cquark\cquarkbar 1}(4430)^\pm$, $P_{c\bar c}(4380)^+$) have been discovered in the past 20 years.  
By unveiling new ways in which QCD forms bound states, these discoveries have sparked 
a highly active area of hadronic physics research, establishing
\textbf{heavy-flavour physics as a unique laboratory for exploring the strong interaction.}

Heavy-flavour research is a major effort in Europe.
Across the continent, there are 72 groups working on the \lhcb experiment, 34 on the
\belletwo experiment, and 18 on the \besiii experiment. In addition,
a significant number of groups work on heavy-flavour physics within  \atlas and
    \cms, and many European groups are leaders in the related theory.
As in other areas of particle physics, data analysis in heavy-flavour physics is 
often performed in small groups at universities and laboratories, involves detailed and specific collaboration with phenomenologists, and can 
take several years to complete due to its high complexity. Therefore, long-term support for both experimental and theoretical groups involved in heavy-flavour research is essential for the success of the field, as well as for training students and postdocs, ensuring the efficient use of investments in research facilities. 
\textbf{A healthy research programme requires support for CERN and 
other global laboratories, 
as well as the individual experimental and theoretical research
groups throughout European universities and institutes.}

In the past, successful heavy-flavour programmes were carried out at the LEP experiments and SLD, the Tevatron experiments, and ARGUS and CLEO. \babar and Belle dominated the field in the first decade of this century, and are still producing results.  
Today, the leading flavour physics experiments are \lhcb, \belletwo and \besiii, with contributions also from \atlas and \cms. 

In beauty-hadron physics, the majority of measurements in the next decade will come from \lhcb and \belletwo.  
\lhcb benefits from large cross-sections for production of all types of beauty hadrons and from precise decay-time measurement that arises from the high boost of the produced particles. 
\belletwo exploits production of $B$-meson pairs with well known kinematics in a nearly hermetic detector, allowing the reconstruction of final states with missing energy and excellent performance for neutrals, such as photons, as well as $\pi^0$ and $\eta$ mesons.  

Similarly, in charm physics, \lhcb and \belletwo will produce the majority of 
high-statistics results. Additionally, the exploitation of \besiii data will also provide unique measurements in the charm sector, leveraging the reconstruction of $D$ decays with missing energy and quantum correlation in $e^+e^-\to D \bar 
D$ processes. Some of these measurements will be key input to \lhcb and \belletwo analyses to precisely measure \CP violation in $B$ decays. In addition to complementarity, this also marks a notable symbiosis among flavour-physics experiments. 

By contrast, in $\tau$ physics, experimental progress will rely primarily on the \belletwo experiment, 
which will significantly advance this sector on both lepton-flavour universality and lepton-flavour violation measurements, and will contribute to tests of the unitarity of the first row of the CKM matrix. 

Each generation of flavour-physics experiments involves significant technological advances and large luminosity increases. 
\lhcb and \belletwo are pushing the boundaries in hadron and  $e^+ e^-$ collisions, respectively. 
They both use state-of-the-art pixel detectors for vertexing and renewed and innovative particle-identification detectors and electronics. 

\lhcb has demonstrated effective heavy-flavour reconstruction at high pile-up with an upgraded detector (Upgrade I) in Run 3 at the LHC to continue the remarkably successful physics programme achieved during Runs 1 and 2 (2010-2018). The newly deployed all-software trigger is central to the real-time reconstruction at the maximum LHC interaction rate, enabling real-time event selections for final analysis with a large increase of signal rates compared to Run 2. The target is to collect by the end of 2033 five times more data than the combined Run 1 and 2 samples. Beyond that, the \lhcb Upgrade~II~\cite{LHCb-UII-FTDR} is proposed to collect five times more data through the lifetime of the high-luminosity LHC (HL-LHC). Current detector performance must be kept at pile-up values of up to about 40, which will require the replacement of all of the existing spectrometer components to increase the granularity, the reduction of the amount of material in the detector, and the development of new technologies with precision timing of the order of a few tens of picoseconds. With Upgrade II, the new physics mass scale probed, for fixed couplings, would almost double as compared with the pre-HL-LHC era. 

The SuperKEKB accelerator and the \belletwo detector are a major upgrade of the KEKB/Belle complex to push the intensity-frontier at colliders to unprecedented luminosities, incidentally providing a test bed for next-generation $e^+e^-$ machines such as FCC-ee or CEPC. SuperKEKB obtained the highest recorded instantaneous luminosity for a collider, $5.1\times 10^{34}$\,cm$^{-2}$\,s$^{-1}$, in December 2024. Although \belletwo has collected a smaller sample than Belle thus far, thanks to a higher data quality and refined analysis techniques, several measurements are already competitive or better than those obtained at previous $B$-factories, demonstrating the excellent capabilities of the experiment. To achieve the design luminosity, a long shut-down after 2028 for an upgrade of the SuperKEKB interaction region, which will require a redesign of the vertex detector, is under discussion. This shut-down also provides an opportunity to upgrade many \belletwo subsystems, to improve the robustness against backgrounds and the radiation resistance, thus providing larger safety factors to run at high luminosity~\cite{BelleII-U-FTDR}. \belletwo ultimately aims at accumulating fifty times more data than \belle by the second half of the next decade, possibly introducing also polarisation of the electron beam, which would add to the current physics reach an unique programme of measurements in the weak sector.  

The heavy flavour measurements that can be carried out at \lhcb Upgrade~II and the upgraded \belletwo are unique and for many of them there are no foreseen facilities in the next many decades that will supersede them. \textbf{Support for the construction and full exploitation of \lhcb Upgrade~II and upgrades of \belletwo is essential to advance the heavy-flavour physics programme and probe the flavour structure of BSM physics with unprecedented precision.}
\textbf{The dual approach of conducting heavy-flavour physics in both the high-statistics hadronic environment and the controlled} $\bm{e^+ e^-}$ \textbf{environment 
is key to realise the full physics potential and should be strongly supported.}

\besiii has recently gone through an upgrade of the inner tracker and will exploit
larger instantaneous luminosities from an upgrade of BEPCII, with also an extension of its centre-of-mass energy up to 2.5\,GeV expected in a few years. After that, a possible super charm-tau
factory that is being studied in China will provide data samples of $\tau$ pairs and quantum-correlated charm-meson pairs about a factor of 100 larger than those of BEPCII~\cite{Achasov:2023gey}. 

Alongside the significant advancements in experimental knowledge that \lhcb, \belletwo, \besiii, and their upgrades will yield, corresponding progress is expected in lattice QCD calculations and the development of other theoretical tools, enhancing the precision of indirect searches for BSM physics.

Heavy-flavour physics will certainly continue to play its important role in the post-LHC era.  
The abundant production of beauty and charm hadrons from about $5\times 10^{12}$ $Z$ decays expected at FCC-ee or CEPC would offer great and complementary opportunities in heavy-flavour physics~\cite{Monteil:2021ith,Ai:2024nmn}.  
A wide range of measurements beyond those unique to \lhcb and \belletwo---including rare decays, \CP-violation studies, and spectroscopy---would benefit from the experimental environment offered by FCC-ee/CEPC. This includes a relatively low background, a high Lorentz boost, fully efficient triggers, and access to the full spectrum of hadron species.
The decays of on-shell $W$ bosons could provide an opportunity to improve some CKM matrix elements, such as the strength of the bottom-to-charm quark coupling~\cite{Marzocca:2024mkc}. In $\tau$ physics, such facilities would have unsurpassed physics reach in almost all measurements.

A full exploitation of flavour physics at FCC-ee/CEPC places specific constraints and challenges on detector design. 
Precise vertexing is essential for all heavy-flavour measurements, playing a particular role in decay-time-dependent measurement, to minimise dilution of \CP asymmetries, or in the reconstruction of semitauonic $B$ decays, to exploit topological constraints in kinematic fits. Excellent energy resolution of electromagnetic calorimeters would leverage the relatively low-multiplicity environment at $e^+e^-$ colliders to make high precision studies involving final states with photons, as well as $\pi^0$ and $\eta$ mesons. Finally, an extensive flavour-physics programme requires high-performance for the identification of pions, kaons, protons and leptons over a very wide momentum range. Maximising particle-identification performances, tracking volume, and energy resolution in a nearly hermetic instrument poses significant design challenges that must be taken into account early on when designing future detectors. 

 \textbf{In general, it is crucial that new facilities account for opportunities to further explore heavy-flavour physics. In this respect, we support a long-term plan that fully integrates flavour physics in an {\boldmath $\epem$} collider that devotes significant collision time to run as a {\boldmath $Z$} factory.} 
 
 For a future circular, very high-energy $pp$ collider, dedicated studies on the integration of flavour-physics experiments should begin early in its design. This machine will present significant challenges, potentially necessitating a dedicated experiment along the lines of \lhcb.

In summary, heavy-flavour physics plays a vital role in advancing the search for BSM physics and deepening our understanding of the SM, a role that will remain essential in the foreseeable future. To fully harness its potential, we strongly advocate for the European Strategy for Particle Physics to support both experimental and theoretical research in this field, ensuring sustained progress in the mid and long term.

\setboolean{inbibliography}{true}
\bibliographystyle{LHCb}
\bibliography{standard}

\end{document}